\documentclass[12pt, reqno] {article}
\usepackage{latexsym}
\usepackage{amsmath}
\usepackage{mathrsfs}
\usepackage{amsfonts}
\usepackage{graphicx}
\usepackage{wrapfig}
\usepackage{xcolor}
\usepackage{cite}
\usepackage[margin=1in]{geometry}
 \usepackage[]{hyperref}
 \hypersetup{colorlinks=true,
citecolor=mdtRed,
linkcolor=gray,
urlcolor=mdtRed
}
\colorlet{mdtRed}{red!50!black}

\usepackage{tikz}
\usetikzlibrary{arrows.meta,decorations.markings,decorations,decorations.text,decorations.pathreplacing}

\tikzset{
  mid arrow/.style={
    postaction={
      decorate,
      decoration={
        markings,
        mark=at position 0.5 with {\arrow{Stealth[length=6pt,width=6pt]}}
      }
    }
  }
}

\usepackage{tikz-3dplot}
\tdplotsetmaincoords{20}{110}
\usetikzlibrary{decorations.pathreplacing,calc}
\usetikzlibrary{decorations.pathmorphing}

\definecolor{blue5}{rgb}{0, 0, 0.66666}
\definecolor{red4}{rgb}{0.8, 0, 0}
\definecolor{navyblue}{rgb}{0, 0.2745, 0.67843}
\definecolor{cornflowerblue}{rgb}{0.388235, 0.6941176, 0.898039}
\usetikzlibrary{decorations.text,decorations.pathmorphing}
\usepackage{subcaption}
\usepackage{setspace}
\newcommand{\bea}{\begin{eqnarray}}
\newcommand{\eea}{\end{eqnarray}}
\newcommand{\be}{\begin{equation}}
\newcommand{\ee}{\end{equation}}
\numberwithin{equation}{section}

\allowdisplaybreaks

\begin{document}
\begin{titlepage}  
\pagestyle{empty}
\baselineskip=21pt
\vspace{10cm}

\begin{center}
{\bf {\large Before the Bang: Wormholes at the Dawn of the Universe}}
\end{center}
\begin{center}
\vskip 0.3in
{\bf Panos Betzios}$^{1}$, {\bf Paul Ghiringhelli}$^{2}$, {\bf Ioannis D.~Gialamas}$^{3}$
and {\bf Olga Papadoulaki}$^{2}$
\vskip 0.1in
{\it
$^1${\href{https://www.ugent.be/we/physics-astronomy/en}{Department of Physics and Astronomy},
Ghent University, \\ Krijgslaan, 281-S9, 9000 Gent, Belgium}\\
$^2${\href{https://www.cpht.polytechnique.fr/?q=en}{CPHT, CNRS, École polytechnique, Institut Polytechnique de Paris}, \\ 91120 Palaiseau, France}\\
$^3${\href{https://kbfi.ee/high-energy-and-computational-physics/?lang=en}{Laboratory of High Energy and Computational Physics}, NICPB, \\
R{\"a}vala pst.~10, Tallinn, 10143, Estonia}}\\
\vspace{5mm}

{\it E-mails}:\\ \href{panos.betzios@ugent.be}{panos.betzios@ugent.be}, 
\href{paul.ghiringhelli@polytechnique.edu}{paul.ghiringhelli@polytechnique.edu}, 
\href{ioannis.gialamas@kbfi.ee}{ioannis.gialamas@kbfi.ee}, \\
\href{olga.papadoulaki@polytechnique.edu}{olga.papadoulaki@polytechnique.edu} (corresponding author)

\vspace{1cm}
{\bf Abstract}
\end{center}
\baselineskip=18pt \noindent  

This essay discusses recent progress on Euclidean wormholes as candidate contributions to the Universe's initial quantum state.
The comparison with the Hartle--Hawking no-boundary proposal highlights both a conceptual affinity and genuine advance: wormholes retain the relevance of Euclidean-saddles as encoders of properties of cosmological wavefunctions, while they broaden the class of regular saddles that are physically relevant for inflating universes and are capable of resolving issues that plague the no-boundary proposal. The principal achievement of the wormhole program is to enlarge the semiclassical initial-condition landscape in a way that is physically rich, conforms with Holographic expectations and as such becomes increasingly relevant for early-universe model building, within UV complete theories of quantum gravity.


\vfill

\begin{center}
{\it Essay written for the Gravity Research Foundation 2026 Awards for Essays on Gravitation} \\
\medskip
Submission date: 30 March 2026
\end{center}

\end{titlepage}
\baselineskip=18pt

\section{Introduction}

\begin{center}
\textit{Before the bang, space could have connected through a narrow tunnel -- a wormhole providing a smooth passage from a Euclidean region to the expanding Universe we observe.}
\end{center}

What fixed the initial conditions of our Universe remains one of the most delicate open problems in quantum gravity. While inflation~\cite{Guth:1980zm,Linde:1981mu} successfully accounts for the observed structure of the late Universe and the near scale-invariance of primordial perturbations~\cite{Starobinsky:1979ty,Mukhanov:1981xt,Hawking:1982cz,Starobinsky:1982ee,Guth:1982ec,Bardeen:1983qw}, it does not by itself explain the origin of the cosmological state from which expansion begins. Progress on the problem of cosmological initial conditions is closely tied to the gravitational path integral, which motivates the search for a well-controlled semiclassical approximation capable of capturing the most important properties of the ``wavefunction of the Universe". Within this framework, the central issue becomes identifying the semi-classical geometries that plausibly dominate the path integral in defining the initial state rather than the subsequent inflationary dynamics.

\begin{figure}[h!]
\centering

\begin{subfigure}{0.48\textwidth}
\centering
\resizebox{!}{2.8cm}{
\begin{tikzpicture}[xscale=0.35, yscale=0.4]
\hspace{0.3cm}

\path[fill=cyan!10, opacity=0.25]
  (-7.1,0)
arc[start angle=180,end angle=270,x radius=1,y radius=3] 
  .. controls (-5.,-2.2) and (-3.,-0.8) .. (-0.5,-0.8)
  .. controls (0.1,-0.8) and (1.4,-2) .. (2.,-2)     
arc[start angle=-90,end angle=-270,x radius=0.6,y radius=2]
  .. controls (1.4,2) and (0.1,0.8) .. (-0.5,0.8)
  .. controls (-3.,0.8) and (-5.,2.2) .. (-6.1,3)
arc[start angle=90,end angle=180,x radius=1,y radius=3]
  -- cycle;

\path[fill=cyan!10, opacity=0.25]
  (10.1,-3)
  .. controls (9.,-2.2) and (7.,-0.8) .. (4.5,-0.8)
  .. controls (3.9,-0.8) and (2.6,-2.) .. (2,-2)
arc[start angle=-90,end angle=-270,x radius=0.6,y radius=2]
  .. controls (2.6,2.) and (3.9,0.8) .. (4.5,0.8)
  .. controls (7.,0.8) and (9.,2.2) .. (10.1,3)
arc[start angle=90,end angle=-90,x radius=1,y radius=3]
  -- cycle;

\draw[thick, cornflowerblue, dotted] (-6.1,0) ellipse (1 and 3);
\draw[thick,  cornflowerblue, dotted] (2,0) ellipse (0.6 and 2);
\draw[thick,  cornflowerblue, dotted] (10.1,0) ellipse (1 and 3);

\draw[thick, navyblue, decorate,
      decoration={
          text along path,
          text={|\scriptsize|EAdS boundary},
          text align=center,
          text color=navyblue,
          reverse path,
          raise=3.5pt,
          text format delimiters={|}{|}
      }
]
(-6.1,3) arc[start angle=90,end angle=270,x radius=1,y radius=3];

\draw[thick, navyblue, decorate,
      decoration={
          text along path,
          text={|\scriptsize|EAdS boundary},
          text align=center,
          text color=navyblue,
          reverse path,
          raise=0.pt,
          text format delimiters={|}{|}
      }
]
(10.8,3) arc[start angle=90,end angle=-90,x radius=1,y radius=3];

\draw[thick, cornflowerblue]
  (-6,3) .. controls (-5.,2.2) and (-3,0.8) .. (-0.5,0.8)
        .. controls (.1,0.8) and (1.4,2.) .. (2,2);

\draw[thick, cornflowerblue]
  (10,3) .. controls (9,2.2) and (7.,0.8) .. (4.5,0.8)
        .. controls (3.9,0.8) and (2.6,2.) .. (2,2);

\draw[thick, cornflowerblue]
  (-6,-3) .. controls (-5.,-2) and (-3,-0.8) .. (-0.5,-0.8)
         .. controls (0.1,-0.8) and (1.4,-2.) ..(2,-2);

\draw[thick, cornflowerblue]
  (10,-3) .. controls (9.,-2.2) and (7.,-0.8) .. (4.5,-0.8)
        .. controls (3.9,-0.8) and (2.6,-2.) .. (2,-2);

\draw[->,  line width=0.9pt,  opacity=0.2] (-6,0) -- (10.2,0);
\node[above] at (9.7,0.) {\scriptsize\color{gray}$\tau$};

\end{tikzpicture}
}
\end{subfigure}
\hfill
\begin{subfigure}{0.4\textwidth}
\centering
\resizebox{!}{2.8cm}{
\begin{tikzpicture}[xscale=0.35, yscale=0.4]

\path[fill=cyan!10, opacity=0.25]
  (-7.1,0)
  arc[start angle=180,end angle=270,x radius=1,y radius=3] 
  .. controls (-5,-2.2) and (-3,-0.8) .. (-0.5,-0.8)
  .. controls (0.1,-0.8) and (1.4,-2) .. (2,-2)     
  arc[start angle=-90,end angle=-270,x radius=0.6,y radius=2]
  .. controls (1.4,1.95) and (0.6,1.4) .. (0,0.8)
  .. controls (-3,1) and (-4.8,2) .. (-6.1,3)
  arc[start angle=90,end angle=180,x radius=1,y radius=3]
  -- cycle;

\path[fill=red, opacity=0.08]
  (1.4,0)
  arc[start angle=180,end angle=-90,x radius=0.6,y radius=2] 
  .. controls (3.5,-2) .. (6.1,-3)                    
  arc[start angle=-90,end angle=90,x radius=1,y radius=3]  
  .. controls (3.5,2) .. (2,2)                        
  arc[start angle=90,end angle=180,x radius=0.6,y radius=2]  
  -- cycle;

\pgfdeclareradialshading{inflationellipse}{\pgfpoint{0cm}{0cm}}{
  color(0cm)=(red4!80);
  color(0.6cm)=(orange!40);
  color(0.9cm)=(yellow!10)
}

\shade[shading=inflationellipse, opacity=0.6] (2,0) ellipse (0.6 and 2);

\draw[thick, navyblue, decorate,
      decoration={
          text along path,
          text={|\scriptsize|EAdS boundary},
          text align=center,
          text color=navyblue,
          reverse path,
          raise=3.5pt,
          text format delimiters={|}{|}
      }
]
(-6.1,3) arc[start angle=90,end angle=270,x radius=1,y radius=3];

\draw[thick, red4, decorate,
      decoration={
          text along path,
          text={|\scriptsize|Inflation},
          text align=center,
          text color=red4,
          reverse path,
          raise=0.pt,
          text format delimiters={|}{|}
      }
]
(2.2,3) arc[start angle=95,end angle=-90,x radius=1,y radius=3];

\draw[thick, red4, decorate,
      decoration={
          text along path,
          text={|\tiny|Lorentzian evolution},
          text align=center,
          text color=red4,
          reverse path,
          raise=0.pt,
          text format delimiters={|}{|}
      }
]
(6.8,3) arc[start angle=90,end angle=-90,x radius=1,y radius=3];

\draw[thick, cornflowerblue, dotted] (-6.1,0) ellipse (1 and 3);
\draw[thick, red4] (6.1,0) ellipse (1 and 3);
\draw[thick, red4, dotted] (2,0) ellipse (0.6 and 2);

\draw[thick, cornflowerblue]
  (-6,3) .. controls (-5.,2.2) and (-3,0.8) .. (-0.5,0.8)
        .. controls (.1,0.8) and (1.4,2.) .. (2,2);

\draw[thick, red4] (2,2) .. controls (3.5,2) .. (6,3);

\draw[thick, cornflowerblue]
  (-6,-3) .. controls (-5.,-2) and (-3,-0.8) .. (-0.5,-0.8)
         .. controls (0.1,-0.8) and (1.4,-2.) .. (2,-2);

\draw[thick, red4] (2,-2) .. controls (3.5,-2) .. (6,-3);

\draw[->,  line width=0.9pt,  opacity=0.2] (-6,0) -- (2,0);
\node[above] at (1.1,0.) {\scriptsize\color{gray}$\tau$};

\draw[->,  line width=0.9pt,  opacity=0.2] (2.2,0) -- (6.6,0);
\node[above] at (6.,0.) {\scriptsize\color{gray}$t$};

\end{tikzpicture}
}
\end{subfigure}

\caption{Schematic illustration of the wineglass wormhole proposal for the birth of the Universe. 
\emph{Left:} The Euclidean wineglass wormhole geometry connecting two asymptotic EAdS boundaries. 
\emph{Right:} Upon analytic continuation ($\tau=it$), one half of the wormhole evolves into a Lorentzian quasi-$dS_4$ spacetime describing an expanding universe.}
\label{fig:wormholes}
\end{figure}
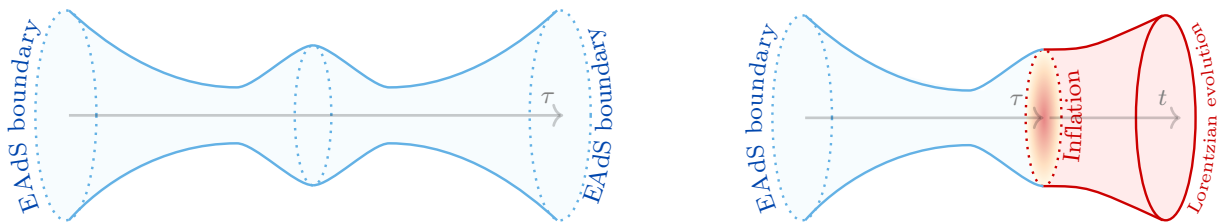

A particularly compelling possibility is provided by Euclidean wormhole solutions that connect an asymptotically Euclidean anti-de Sitter (AdS) region to a Lorentzian expanding universe~\cite{Betzios:2024oli,Betzios:2024zhf,Betzios:2026rbv} (see figure~\ref{fig:wormholes}).
In this framework, the onset of cosmological evolution is determined dynamically by a geometric saddle of the Euclidean gravitational equations, rather than being imposed through an abstract and ad-hoc initial condition.
Importantly for the class of solutions considered here, the relevant saddles are not compact geometries without boundary as in the traditional no-boundary proposal by Hartle and Hawking~\cite{Hartle:1983ai}; instead, they possess well-defined asymptotic ($AdS$) boundaries equipped with a non trivial Hilbert space of states, consistent with Holographic expectations. This feature gives the construction of a more solid basis within the gravitational path integral~\cite{Betzios:2026rbv} and allows for a more controlled semiclassical interpretation.


Within this framework, the wave function of the Universe is approximated semiclassically by saddle points of the gravitational path integral
where each saddle is potentially subject to boundary conditions (should boundaries exist) and where the paths are in general defined in the complexified field space. The Hartle--Hawking proposal chooses compact, regular ``no-boundary'' Euclidean saddles in the far past, leading to an initial wavefunction
\be\label{Noboundarywvf}
 \Psi_{N.B.}[h_{i j}, \phi_0] = \int^{g_{i j}(0) = h_{i j}} \mathcal{D}g \int^{\phi(0) = \phi_0} \mathcal{D} \phi \,e^{- S_E[g_{\mu \nu}, \phi]} \, ,
\ee
and to complexified paths and saddles, if one chooses to evolve the wavefunction in the real Lorentzian time $t$.

On the other hand, the more recent wormhole proposal posits that other regular but topologically nontrivial saddles should contribute
in determining the wavefunction of the Universe. In particular, the initial ($t=0$) wavefunction is given by the following Euclidean path integral
\be
\langle g^{\partial}_{i j}, J^{\partial}_\phi | h_{i j}, \phi_0 \rangle \, \equiv \, \Psi[ g^{\partial}_{i j}, J^{\partial}_\phi \, ; \, h_{i j}, \phi_0] = \int^{g_{i j}(0) = h_{i j}}_{g_{i j}(\partial \mathcal{M}) \rightarrow g_{i j}^{\partial}} \mathcal{D}g \int^{\phi(0) = \phi_0}_{\phi(\partial \mathcal{M}) \rightarrow J^{\partial}_\phi} \mathcal{D} \phi \, e^{- S_E[g_{\mu \nu}, \phi]} \, ,
\ee
where  the initial wavefunction of the Universe depends now on asymptotically AdS boundary conditions $g^{\partial}_{i j}, J^{\partial}_\phi $ of the various fields.

In this formulation~\cite{Abdalla:2026mxn}, the probability for the Universe to emerge in a specific state is computed by considering the relative weights of all contributing saddles. Specifically, the probability $P$ of transitioning from the asymptotic boundary ($AB$) state to various interior boundary ($IB$) states is given by the ratio of the relevant transition amplitudes over the sum of all regular configurations in the path integral. While the transition amplitudes themselves comprise of only connected configurations, the denominator naturally contains all possible regular saddles, rendering any such probability $P<1$. In particular, in certain examples, one finds both the standard wormhole geometries ($\vcenter{\hbox{\resizebox{!}{0.3cm}{
\begin{tikzpicture}[xscale=0.5, yscale=0.4, baseline=(current bounding box.center)]
\path[fill=cyan!10, opacity=0.25] (-7.2,4.5) .. controls (-4.5,1.5) and (-2,1.2) .. (0,1.2) .. controls (2,1.2) and (4.5,1.5) .. (7.2,4.5) arc[start angle=90, end angle=-90, x radius=1.2, y radius=4.5] .. controls (4.5,-1.5) and (2,-1.2) .. (0,-1.2) .. controls (-2,-1.2) and (-4.5,-1.5) .. (-7.2,-4.5) arc[start angle=270, end angle=90, x radius=1.2, y radius=4.5] -- cycle;
\draw[line width=1pt, cornflowerblue, dotted] (-7.2,0) ellipse (1.2 and 4.5);
\draw[line width=1pt, cornflowerblue, dotted] (7.2,0) ellipse (1.2 and 4.5);
\draw[line width=1pt, cornflowerblue, dotted] (0,0) ellipse (0.35 and 1.2);
\draw[line width=1pt, cornflowerblue] (-7,4.5) .. controls (-4.5,1.5) and (-2,1.2) .. (0,1.2) .. controls (2,1.2) and (4.5,1.5) .. (7,4.5);
\draw[line width=1pt, cornflowerblue] (-7,-4.5) .. controls (-4.5,-1.5) and (-2,-1.2) .. (0,-1.2) .. controls (2,-1.2) and (4.5,-1.5) .. (7,-4.5);
\end{tikzpicture}
}}}$), the disconnected configurations ($\vcenter{\hbox{\resizebox{!}{0.3cm}{
\begin{tikzpicture}[xscale=0.5, yscale=0.4, baseline=(current bounding box.center)]
\path[fill=cyan!10, opacity=0.25] (-7.4,4.5) .. controls (-4.5,1.5) and (-0.5,1.5) .. (-0.5,0) .. controls (-0.5,-1.5) and (-4.5,-1.5) .. (-7.4,-4.5) arc[start angle=270, end angle=90, x radius=1.2, y radius=4.5] -- cycle;
\path[fill=cyan!10, opacity=0.25] (7.4,4.5) .. controls (4.5,1.5) and (0.5,1.5) .. (0.5,0) .. controls (0.5,-1.5) and (4.5,-1.5) .. (7.4,-4.5) arc[start angle=-90, end angle=90, x radius=1.2, y radius=4.5] -- cycle;
\draw[line width=1pt, cornflowerblue] (-7.2,4.5) .. controls (-4.5,1.5) and (-0.5,1.5) .. (-0.5,0) .. controls (-0.5,-1.5) and (-4.5,-1.5) .. (-7.2,-4.5);
\draw[line width=1pt, cornflowerblue] (7.2,4.5) .. controls (4.5,1.5) and (0.5,1.5) .. (0.5,0) .. controls (0.5,-1.5) and (4.5,-1.5) .. (7.2,-4.5);
\draw[line width=1pt, cornflowerblue, dotted] (-7.4,0) ellipse (1.2 and 4.5);
\draw[line width=1pt, cornflowerblue, dotted] (7.4,0) ellipse (1.2 and 4.5);
\end{tikzpicture}
}}}$), the recently proposed wineglass wormholes ($\vcenter{\hbox{\resizebox{!}{0.3cm}{
\begin{tikzpicture}[xscale=0.35, yscale=0.4, baseline=(current bounding box.center)]
\path[fill=cyan!10, opacity=0.25] (-7.1,0) arc[start angle=180,end angle=270,x radius=1,y radius=3] .. controls (-5.,-2.2) and (-3.,-0.8) .. (-0.5,-0.8) .. controls (0.1,-0.8) and (1.4,-2) .. (2.,-2) arc[start angle=-90,end angle=-270,x radius=0.6,y radius=2] .. controls (1.4,2) and (0.1,0.8) .. (-0.5,0.8) .. controls (-3.,0.8) and (-5.,2.2) .. (-6.1,3) arc[start angle=90,end angle=180,x radius=1,y radius=3] -- cycle;
\path[fill=cyan!10, opacity=0.25] (10.1,-3) .. controls (9.,-2.2) and (7.,-0.8) .. (4.5,-0.8) .. controls (3.9,-0.8) and (2.6,-2.) .. (2,-2) arc[start angle=-90,end angle=-270,x radius=0.6,y radius=2] .. controls (2.6,2.) and (3.9,0.8) .. (4.5,0.8) .. controls (7.,0.8) and (9.,2.2) .. (10.1,3) arc[start angle=90,end angle=-90,x radius=1,y radius=3] -- cycle;
\draw[thick, cornflowerblue, dotted] (-6.1,0) ellipse (1 and 3);
\draw[thick, cornflowerblue, dotted] (2,0) ellipse (0.6 and 2);
\draw[thick, cornflowerblue, dotted] (10.1,0) ellipse (1 and 3);
\draw[thick, cornflowerblue] (-6,3) .. controls (-5.,2.2) and (-3,0.8) .. (-0.5,0.8) .. controls (.1,0.8) and (1.4,2.) .. (2,2);
\draw[thick, cornflowerblue] (10,3) .. controls (9,2.2) and (7.,0.8) .. (4.5,0.8) .. controls (3.9,0.8) and (2.6,2.) .. (2,2);
\draw[thick, cornflowerblue] (-6,-3) .. controls (-5.,-2) and (-3,-0.8) .. (-0.5,-0.8) .. controls (0.1,-0.8) and (1.4,-2.) ..(2,-2);
\draw[thick, cornflowerblue] (10,-3) .. controls (9.,-2.2) and (7.,-0.8) .. (4.5,-0.8) .. controls (3.9,-0.8) and (2.6,-2.) .. (2,-2);
\end{tikzpicture}
}}}$), as well as oscillatory ones to contribute~\cite{Betzios:2026rbv}. The normalised probability is then expressed by the following relationship
\begin{align}
    P(\text{\small $AB$} \rightarrow \text{\small $IB_{1,2}$}) = \frac{\left|\, 
    \text{\tiny $AB$} \!\, \vcenter{\hbox{\resizebox{!}{0.4cm}{
        \begin{tikzpicture}[xscale=0.5, yscale=0.4, baseline=(current bounding box.center)]
            \path[fill=cyan!10, opacity=0.25] (-7.4,4.5) .. controls (-4.5,1.5) and (-2,1.2) .. (-0.5,1.2) arc[start angle=90, end angle=-90, x radius=0.35, y radius=1.2] .. controls (-2,-1.2) and (-4.5,-1.5) .. (-7.4,-4.5) arc[start angle=270, end angle=90, x radius=1.2, y radius=4.5] -- cycle;
            \draw[line width=1pt, cornflowerblue] (-7.2,4.5) .. controls (-4.5,1.5) and (-2,1.2) .. (-0.5,1.2);
            \draw[line width=1pt, cornflowerblue] (-7.2,-4.5) .. controls (-4.5,-1.5) and (-2,-1.2) .. (-0.5,-1.2);
            \draw[line width=1pt, cornflowerblue, dotted] (-7.4,0) ellipse (1.2 and 4.5);
            \draw[line width=1pt, cornflowerblue, dotted] (-0.5,0) ellipse (0.35 and 1.2);
        \end{tikzpicture}
    }}} \!\, \text{\tiny $IB_1$} 
    \quad\text{or}\quad 
    \text{\tiny $AB$} \!\, \vcenter{\hbox{\resizebox{!}{0.4cm}{
        \begin{tikzpicture}[xscale=0.35, yscale=0.4, baseline=(current bounding box.center)]
            \path[fill=cyan!10, opacity=0.25] (-7.1,0) arc[start angle=180,end angle=270,x radius=1,y radius=3] .. controls (-5.,-2.2) and (-3.,-0.8) .. (-0.5,-0.8) .. controls (0.1,-0.8) and (1.4,-2) .. (2,-2) arc[start angle=-90, end angle=90, x radius=0.6, y radius=2] .. controls (1.4,2) and (0.1,0.8) .. (-0.5,0.8) .. controls (-3.,0.8) and (-5.,2.2) .. (-6.1,3) arc[start angle=90,end angle=180,x radius=1,y radius=3] -- cycle;
            \draw[thick, cornflowerblue] (-6,3) .. controls (-5.,2.2) and (-3,0.8) .. (-0.5,0.8) .. controls (.1,0.8) and (1.4,2.) .. (2,2);
            \draw[thick, cornflowerblue] (-6,-3) .. controls (-5.,-2) and (-3,-0.8) .. (-0.5,-0.8) .. controls (0.1,-0.8) and (1.4,-2.) ..(2,-2);
            \draw[thick, cornflowerblue, dotted] (-6.1,0) ellipse (1 and 3);
            \draw[thick, cornflowerblue, dotted] (2,0) ellipse (0.6 and 2);
        \end{tikzpicture}
    }}} \!\, \text{\tiny $IB_2$} \quad \cdots \right|^2}{
    \left( \text{\tiny $AB$} \!\, \vcenter{\hbox{\resizebox{!}{0.38cm}{
        \begin{tikzpicture}[xscale=0.5, yscale=0.4, baseline=(current bounding box.center)]
            \path[fill=cyan!10, opacity=0.25] (-7.4,4.5) .. controls (-4.5,1.5) and (-0.5,1.5) .. (-0.5,0) .. controls (-0.5,-1.5) and (-4.5,-1.5) .. (-7.4,-4.5) arc[start angle=270, end angle=90, x radius=1.2, y radius=4.5] -- cycle;
            \path[fill=cyan!10, opacity=0.25] (7.4,4.5) .. controls (4.5,1.5) and (0.5,1.5) .. (0.5,0) .. controls (0.5,-1.5) and (4.5,-1.5) .. (7.4,-4.5) arc[start angle=-90, end angle=90, x radius=1.2, y radius=4.5] -- cycle;
            \draw[line width=1pt, cornflowerblue] (-7.2,4.5) .. controls (-4.5,1.5) and (-0.5,1.5) .. (-0.5,0) .. controls (-0.5,-1.5) and (-4.5,-1.5) .. (-7.2,-4.5);
            \draw[line width=1pt, cornflowerblue] (7.2,4.5) .. controls (4.5,1.5) and (0.5,1.5) .. (0.5,0) .. controls (0.5,-1.5) and (4.5,-1.5) .. (7.2,-4.5);
            \draw[line width=1pt, cornflowerblue, dotted] (-7.4,0) ellipse (1.2 and 4.5);
            \draw[line width=1pt, cornflowerblue, dotted] (7.4,0) ellipse (1.2 and 4.5);
        \end{tikzpicture}
    }}} \!\, \text{\tiny $AB$} +
    \text{\tiny $AB$} \!\, \vcenter{\hbox{\resizebox{!}{0.38cm}{
        \begin{tikzpicture}[xscale=0.5, yscale=0.4, baseline=(current bounding box.center)]
            \path[fill=cyan!10, opacity=0.25] (-7.2,4.5) .. controls (-4.5,1.5) and (-2,1.2) .. (0,1.2) .. controls (2,1.2) and (4.5,1.5) .. (7.2,4.5) arc[start angle=90, end angle=-90, x radius=1.2, y radius=4.5] .. controls (4.5,-1.5) and (2,-1.2) .. (0,-1.2) .. controls (-2,-1.2) and (-4.5,-1.5) .. (-7.2,-4.5) arc[start angle=270, end angle=90, x radius=1.2, y radius=4.5] -- cycle;
            \draw[line width=1pt, cornflowerblue, dotted] (-7.2,0) ellipse (1.2 and 4.5);
            \draw[line width=1pt, cornflowerblue, dotted] (7.2,0) ellipse (1.2 and 4.5);
            \draw[line width=1pt, cornflowerblue, dotted] (0,0) ellipse (0.35 and 1.2);
            \draw[line width=1pt, cornflowerblue] (-7,4.5) .. controls (-4.5,1.5) and (-2,1.2) .. (0,1.2) .. controls (2,1.2) and (4.5,1.5) .. (7,4.5);
            \draw[line width=1pt, cornflowerblue] (-7,-4.5) .. controls (-4.5,-1.5) and (-2,-1.2) .. (0,-1.2) .. controls (2,-1.2) and (4.5,-1.5) .. (7,-4.5);
        \end{tikzpicture}
    }}} \!\, \text{\tiny $AB$} +
    \text{\tiny $AB$} \!\, \vcenter{\hbox{\resizebox{!}{0.38cm}{
        \begin{tikzpicture}[xscale=0.35, yscale=0.4, baseline=(current bounding box.center)]
            \path[fill=cyan!10, opacity=0.25] (-7.1,0) arc[start angle=180,end angle=270,x radius=1,y radius=3] .. controls (-5.,-2.2) and (-3.,-0.8) .. (-0.5,-0.8) .. controls (0.1,-0.8) and (1.4,-2) .. (2.,-2) arc[start angle=-90,end angle=-270,x radius=0.6,y radius=2] .. controls (1.4,2) and (0.1,0.8) .. (-0.5,0.8) .. controls (-3.,0.8) and (-5.,2.2) .. (-6.1,3) arc[start angle=90,end angle=180,x radius=1,y radius=3] -- cycle;
            \path[fill=cyan!10, opacity=0.25] (10.1,-3) .. controls (9.,-2.2) and (7.,-0.8) .. (4.5,-0.8) .. controls (3.9,-0.8) and (2.6,-2.) .. (2,-2) arc[start angle=-90,end angle=-270,x radius=0.6,y radius=2] .. controls (2.6,2.) and (3.9,0.8) .. (4.5,0.8) .. controls (7.,0.8) and (9.,2.2) .. (10.1,3) arc[start angle=90,end angle=-90,x radius=1,y radius=3] -- cycle;
            \draw[thick, cornflowerblue, dotted] (-6.1,0) ellipse (1 and 3);
            \draw[thick, cornflowerblue, dotted] (2,0) ellipse (0.6 and 2);
            \draw[thick, cornflowerblue, dotted] (10.1,0) ellipse (1 and 3);
            \draw[thick, cornflowerblue] (-6,3) .. controls (-5.,2.2) and (-3,0.8) .. (-0.5,0.8) .. controls (.1,0.8) and (1.4,2.) .. (2,2);
            \draw[thick, cornflowerblue] (10,3) .. controls (9,2.2) and (7.,0.8) .. (4.5,0.8) .. controls (3.9,0.8) and (2.6,2.) .. (2,2);
            \draw[thick, cornflowerblue] (-6,-3) .. controls (-5.,-2) and (-3,-0.8) .. (-0.5,-0.8) .. controls (0.1,-0.8) and (1.4,-2.) ..(2,-2);
            \draw[thick, cornflowerblue] (10,-3) .. controls (9.,-2.2) and (7.,-0.8) .. (4.5,-0.8) .. controls (3.9,-0.8) and (2.6,-2.) .. (2,-2);
        \end{tikzpicture}
    }}} \!\, \text{\tiny $AB$} + \cdots \right) \langle \text{\small $IB_{1,2}$} | \text{\small $IB_{1,2}$} \rangle}\,.
\end{align}
The resulting probability distribution is therefore sensitive to the topology of the underlying manifold and the specific boundary conditions imposed at the asymptotic region. Such configurations have recently gained significant attention as they provide a bridge between holographic duality and the global structure of spacetime in cosmology~\cite{Betzios:2017krj,Hebecker:2018ofv,Fu:2019oyc,VanRaamsdonk:2021qgv,Antonini:2022blk,Antonini:2022ptt,Jonas:2023ipa,Aguilar-Gutierrez:2023ril,Antonini:2024bbm,Maloney:2025tnn,Lavrelashvili:2026zsw}.
\vspace{0.2cm}

This proposal  delivers a compelling message: 
\textit{Euclidean wormhole saddles are viable structures in quantum cosmology, and they open a broader conceptual and computational arena for understanding cosmological initial conditions.}

\section{Technical core}

The dynamics of a scalar field $\phi$ minimally coupled to gravity, together with additional matter components such as radiation and axion fields, can be described by the action (here in Euclidean signature)
\begin{equation}
\mathcal{S}_E = \int {\rm d}^4x \sqrt{g_E} 
\left(
-\frac{M_P^2}{2} R
+ \frac{1}{2} (\partial_\mu \phi)^2
+ V(\phi)
+ \mathcal{L}^E_i
\right)-M_P^2\int {\rm d}^3x \sqrt{h}K, \qquad i = {\rm rad},\,{\rm axion}\,.
\end{equation}
Assuming a homogeneous and isotropic Universe, with metric in the Euclidean/Lorentzian section
\be
{\rm d}s^2_E = {\rm d}\tau^2 + a^2(\tau){\rm d}\Omega_3^2\, , \qquad \tau = i t \, , \qquad {\rm d}s^2_L = - {\rm d} t^2 + a^2(t){\rm d}\Omega_3^2 \, ,
\ee
and fields depending only on Euclidean time, the equations of motion reduce to
\begin{subequations}
\begin{align}
\label{eq:eqom1}
\frac{a'^2}{a^2} - \frac{1}{a^2}
+ \frac{1}{3M_P^2} \left(V(\phi) - \frac{\phi'^2}{2}\right)
- \frac{\tilde{\rho}_i^E}{a^{n_i}} &= 0\,,
\\
\phi'' + 3 \frac{a'\phi'}{a} - \partial_\phi V(\phi) &= 0\,,
\end{align}
\end{subequations}
where $n_{\rm rad}=4$ and $n_{\rm axion}=6$. The effective densities are defined as $\tilde{\rho}_{\rm rad} = a^4 \rho_{\rm rad}^E / (3 M_P^2)$ and $\tilde{\rho}_{\rm axion} = - q^2 / (6 M_P^2 f_a^2)$, with $\rho_{\rm rad}^E$ being the Euclidean radiation energy density, $q$ a constant axion charge ($H_{ijk}=q \epsilon_{ijk}$), and $f_a$ the axion decay constant. For the axion sector, we have assumed the standard Euclidean Lagrangian $\mathcal{L}_{\rm axion}^E = H_{\mu\nu\rho} H^{\mu\nu\rho} / (12 f_a^2)$, where $H_{\mu\nu\rho}$ is the three-form field strength. In special cases with enhanced symmetries, such as in a $U(1)^3$ model for the radiation, completely analytic solutions can be obtained~\cite{Marolf:2021kjc,Betzios:2026rbv}, though the details lie beyond the scope of this essay.

From equation~\eqref{eq:eqom1}, the influence of each component on the evolution of the universe becomes apparent. Gravity naturally tends to shrink the scale factor, contributing with a curvature term $-1/a^2$. Axions and magnetic fluxes, on the other hand, generate a repulsive effect when the Euclidean energy density is negative\footnote{Interestingly, a negative radiation energy density arises naturally in the Euclidean framework; for instance, in a simple $U(1)$ model the Euclidean energy density is $\rho^E_{\rm rad} = \frac{1}{2} (E^2 - B^2)$, which becomes negative if the magnetic component dominates.}, giving rise to the wormhole neck and facilitating the subsequent accelerated expansion of the universe upon continuing $\tau = i t$.

The Euclidean “wineglass” wormhole solutions of interest are characterized by their asymptotic approach to Euclidean AdS (EAdS) space. Furthermore, we require these solutions to satisfy specific boundary conditions at the $\mathbb{Z}_2$ symmetry center ($\tau=0$), which marks the transition to a Lorentzian manifold:
\begin{align}
\text{\color{cornflowerblue}Euclidean:}&\quad a''(0) < 0\,, \quad a'(0) =0\,,  \quad a(0) =a_{\max}\,, \quad \phi'(0)=0\,, \nonumber
\\[-0.2cm]
& \hspace{3.85cm}\Downarrow \tau = it 
\\[-0.2cm]
\text{\color{red4}Lorentzian:}&\quad \,\,\underbrace{\ddot a(0) > 0}_{\substack{\text{accelerating}\\ \text{expansion}}}\,, \quad \,\dot a(0)=0\,, \quad a(0)=a_{\min}\,, \quad \,\,\underbrace{\dot\phi(0)=0}_{\substack{\text{scalar field}\\ \text{in slow-roll}}}\,. \nonumber
\end{align}
These constraints, which provide the ideal physical conditions for the nucleation of our Universe, are naturally realized by wineglass wormhole geometries~\cite{Betzios:2024oli}. In contrast to simple wormholes that yield crunching cosmologies upon analytic continuation, wineglass solutions inherently give rise to an expanding universe. Furthermore, under appropriate  conditions, they provide the dominant connected contributions to the gravitational path integral, rendering the wineglass proposal not just a merely plausible scenario invoked by an anthropic principle, but a dynamically and physically preferred possibility~\cite{Betzios:2026rbv}.

The models studied in the works~\cite{Betzios:2024zhf,Betzios:2025eev,Betzios:2026rbv} are built on a common architecture that is part of the Standard Model coupled to General Relativity (GR), but the construction of such wormhole solutions is also easily generalizable to theories beyond it~\cite{Cicoli:2023opf}.

The construction of Euclidean solutions is performed for quite general gravitational theories in the presence of additional sectors of scalar and radiation fields. Depending on the model, some of these sectors could correspond to axion-like, or Higgs-like fields, or other generalized dark-sector extensions. The recurrent point is that the interplay of scalar and radiation dynamics can support regular geometries with a throat-like structure, rather than only capped compact geometries in the Euclidean past, such as the one of the no-boundary proposal.

Additionally, the analysis is performed in a semiclassical saddle-point treatment of the path integral that can be applied to any physical model. In this setting, the physical relevance of any wormhole is controlled by at least three ingredients: the existence of a regular solution, a prescription for the integration contour in complexified field/metric space, and a stability analysis of fluctuations around the saddle.

Moreover, the connection of Euclidean configurations to Lorentzian cosmological histories is performed via analytic continuation and smooth matching conditions. This is where the cosmological interpretation enters: a Euclidean saddle is not by itself a model of the Universe unless it yields a viable classical Lorentzian branch, that can expand in a way that conforms with experimental observations.

Taken together, these analyses suggest that wormhole saddles are more than formal curiosities. At the minimum, they appear as semiclassical contributions in the Standard Model coupled to General Relativity and their significance is clear: once wormhole channels are admitted, the architecture of the initial-state for the resulting cosmology becomes richer.

\section{Comparison with the Hartle--Hawking  proposal}
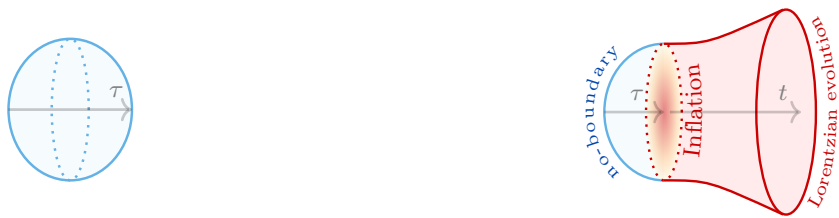
\begin{figure}[h!]
\centering

\begin{subfigure}{0.48\textwidth}
\centering
\resizebox{!}{2.8cm}{
\begin{tikzpicture}[xscale=0.35, yscale=0.4]

\path (-2,-3) (2,3); 

\path[fill=cyan!10, opacity=0.3] (0,0) circle (2);

\draw[thick, cornflowerblue] (0,0) circle (2);
\draw[thick, cornflowerblue, dotted] (0,0) ellipse (0.6 and 2);
\draw[->, line width=0.9pt, opacity=0.2] (-2,0) -- (2,0);
\node[above] at (1.5,0.) {\scriptsize\color{gray}$\tau$};

\end{tikzpicture}
}
\end{subfigure}
\hfill
\begin{subfigure}{0.48\textwidth}
\centering
\resizebox{!}{2.8cm}{
\begin{tikzpicture}[xscale=0.35, yscale=0.4]

\path[fill=cyan!10, opacity=0.3] 
    (2,2) arc[start angle=90, end angle=270, x radius=2, y radius=2]
    -- (2,-2) -- cycle;

\draw[thick, cornflowerblue] (2,2) arc[start angle=90, end angle=270, x radius=2, y radius=2];
\draw[thick, navyblue, decorate,
     decoration={
         text along path,
         text={|\tiny|no-boundary},
         text align=center,
         text color=navyblue,
         reverse path,
         raise=3.5pt,
         text format delimiters={|}{|}
     }
]
(2,2) arc[start angle=90,end angle=270,x radius=2,y radius=2];

\path[fill=red, opacity=0.08]
 (1.4,0)
 arc[start angle=180,end angle=-90,x radius=0.6,y radius=2] 
 .. controls (3.5,-2) .. (6.1,-3)                    
 arc[start angle=-90,end angle=90,x radius=1,y radius=3]  
 .. controls (3.5,2) .. (2,2)                        
 arc[start angle=90,end angle=180,x radius=0.6,y radius=2]  
 -- cycle;

\shade[shading=inflationellipse, opacity=0.6] (2,0) ellipse (0.6 and 2);

\draw[thick, red4] (6.1,0) ellipse (1 and 3);
\draw[thick, red4, dotted] (2,0) ellipse (0.6 and 2);
\draw[thick, red4] (2,2) .. controls (3.5,2) .. (6,3);
\draw[thick, red4] (2,-2) .. controls (3.5,-2) .. (6,-3);

\draw[thick, red4, decorate,
     decoration={
         text along path,
         text={|\scriptsize|Inflation},
         text align=center,
         text color=red4,
         reverse path,
         raise=0.pt,
         text format delimiters={|}{|}
     }
]
(2.2,3) arc[start angle=95,end angle=-90,x radius=1,y radius=3];

\draw[thick, red4, decorate,
     decoration={
         text along path,
         text={|\tiny|Lorentzian evolution},
         text align=center,
         text color=red4,
         reverse path,
         raise=0.pt,
         text format delimiters={|}{|}
     }
]
(6.8,3) arc[start angle=90,end angle=-90,x radius=1,y radius=3];

\draw[->,  line width=0.9pt,  opacity=0.2] (0,0) -- (2,0);
\node[above] at (1.1,0.) {\scriptsize\color{gray}$\tau$};
\draw[->,  line width=0.9pt,  opacity=0.2] (2.2,0) -- (6.6,0);
\node[above] at (6.,0.) {\scriptsize\color{gray}$t$};

\end{tikzpicture}
}
\end{subfigure}

\caption{Schematic illustration of the no-boundary proposal for the birth of the Universe. 
\emph{Left:} The $S^4$ geometry with no boundaries. 
\emph{Right:} Upon analytic continuation ($\tau=it$), a hemisphere evolves into a Lorentzian quasi-$dS_4$ spacetime describing an expanding universe.}
\label{fig:figure2}
\end{figure}

\textbf{Geometric principle:} One could have expected that it is impossible to understand the wavefunction of our Universe using geometric saddles due to the Lorentzian Big-Bang singularity and the breakdown of GR. Nevertheless the no-boundary proposal emphasizes that an excursion to Euclidean signature at early times can avoid this issue, while at the same time a smooth and compact semiclassical geometric saddle (see fig.~\ref{fig:figure2}) can correctly describe the semi-classical (WKB) limit of the wavefunction.  Wormhole approaches preserve this geometric principle but allow additional topological classes of regular saddles and the presence of asymptotic AdS boundaries. In this sense, wormholes are best viewed as an enrichment of the geometric principle.

\textbf{Hilbert space interpretation:}
The no-boundary proposal is considered to be appealing because it restricts to a saddle class for which regularity is the only condition. Various arguments lead to the conclusion that the resulting Hilbert space of states for the bulk quantum gravity theory is restricted to be one dimensional~\cite{McNamara:2020uza,Abdalla:2026mxn}. The wormhole program on the contrary trades that economy for a different saddle inventory, that depends on boundary conditions on the EAdS boundary. The gain is substantial - it gives access to a non-trivial Hilbert space of states that the no-boundary proposal misses.

\textbf{Dependence on contour choice:}
Both frameworks face ambiguities regarding contour choices in complexified path integrals. In the wormhole setting, this is best seen as an opportunity: with a larger and better-structured saddle space, contour principles can be sharpened against a richer set of physically meaningful solutions. Moreover, the EAdS asymptotics lend itself to a definition of the gravitational path integral using a well defined Holographic dual field theory, evading the issue of rigorously defining the gravitational path integral directly in the bulk, and giving clues on the regime of its breakdown and on what replaces the effective gravitational description.

\textbf{Connection to observations:}
No-boundary and wormhole scenarios both seek implications for inflationary priors and cosmological perturbation statistics. In the standard no-boundary proposal the probability for nucleation scales as $P \propto \exp\left[24\pi^2/V(\phi_\star)\right]$, which favors a short duration of inflation and leads to tension with observations.
The wormhole program avoids this problem and strengthens the bridge from formal quantum cosmology to phenomenology by showing concrete pathways through which initial-state weighting can affect inflationary priors.

\section{Merits of the wormhole proposal}

The central novelty is structural. In the standard no-boundary proposal, attention is focused on compact, single-capped regular geometry. The wormhole program expands this to other regular saddles with nontrivial topology and, effectively, to additional channels in configuration space. This enlargement is decisive because probabilities of different Histories of the Universe are weighted by the Euclidean action. If multiple regular saddle families are admitted, relative probabilities for inflationary initial data can be different compared with those of restricted no-boundary saddles, leading to a more coherent cosmological evolution consistent with experimental data.

A second novelty is that inflation becomes a natural and preferred consequence of the early Universe dynamics, when certain classes of wormholes (wineglass) dominate the gravitational path integral. This result is not imposed upon using anthropic or other similar a-posteriori principles, but is dependent on the specific model of our Universe and its precise dynamics, giving the possibility to investigate and constrain the space of low energy effective models.  

A third novelty is model proximity to the Standard Model. Constructions using Higgs inflation models suggest the resulting wormhole cosmology fits within known particle physics~\cite{Betzios:2024zhf}. This improves interpretability and connects initial condition studies to broader field theoretic constraints. Specifically, this scenario is realized if the Standard Model Higgs effective potential develops a second true AdS vacuum at extremely high energies~\cite{Buttazzo:2013uya}. This feature emerges naturally since the running Higgs self coupling can turn the potential negative, a behavior completely consistent with measured Higgs and top quark masses. Within this framework the universe can nucleate from this deep vacuum. Simultaneously the very same Higgs field acts as the inflaton and drives a highly successful and observationally viable period of cosmic inflation~\cite{Bezrukov:2007ep}.

\section{Outlook: Toward a unified initial-state program}

The most constructive way forward is to treat no-boundary and wormhole saddles within a single unified Euclidean framework. In that program, one computes and compares all regular semiclassical channels allowed by different theories/models of the early Universe, and then asks which channels survive contour, stability, and phenomenological consistency tests.

Concretely, progress would come from: (i) sharper contour principles tied to physical boundary data; (ii) uniform fluctuation analyses across saddle families; (iii) model-to-observation pipelines that translate modified initial-state weights into CMB- and large-scale-structure-level signatures. If this agenda succeeds, the community may move from qualitative statements about possible origins to quantitative Bayesian comparison and contrast of various initial-state proposals and models with observational data.

The central point is that wormhole contributions now belong in the foreground of serious initial-state inference, where they can guide both theoretical development and next-generation cosmological model comparison.

\section{Conclusion}
The proposal laid out in the recent works~\cite{Betzios:2024oli,Betzios:2024zhf,Betzios:2026rbv} establishes a powerful claim: Euclidean wormholes are robust semiclassical contributions to cosmological initial conditions in scalar-radiation-gravity theories. Their principal value is to enlarge the regular-saddle landscape beyond the no-boundary geometries and to elevate initial-state cosmology into a genuinely multi-channel theory of origin. Analyses of fluctuations around wormhole saddles can then deliver stronger selection criteria helping to identify which geometric channels are consistent with late time phenomenology.

Relative to Hartle--Hawking, the wormhole proposal  offers broader structural possibilities, richer correlations and constraints between different sectors of the Standard Model coupled to GR, and a more ambitious and microscopic explanatory horizon for the physics of the early Universe. The most compelling interpretation is therefore plural and dynamic: the physics of the early Universe is best captured by a weighted ensemble of regular Euclidean saddles, with certain wormhole channels playing a central role in the genesis of a quantum-cosmological theory that is completely consistent with observations, while at the same time having the potential to microscopically describe the origins of the Universe.

\section*{Acknowledgements}
PB acknowledges financial support from the European Research Council (grant BHHQG-101040024), funded by the European Union.
\\
The work of IDG was supported by the Estonian Research Council grants PSG1132, TARISTU24-TK10, TARISTU24-TK3, and the CoE program TK202 ``Foundations of the Universe'’. \\
Views and opinions expressed are those of the authors only and do not necessarily reflect those of the European Union or the European Research Council. Neither the European Union nor the granting authority can be held responsible for them.

\bibliographystyle{JHEP}
\bibliography{biblio}

\providecommand{\noopsort}[1]{}\providecommand{\singleletter}[1]{#1}%
\providecommand{\href}[2]{#2}\begingroup\raggedright\begin{thebibliography}{10}

\bibitem{Guth:1980zm}
A.~H. Guth, \emph{{The Inflationary Universe: A Possible Solution to the Horizon and Flatness Problems}}, \href{http://dx.doi.org/10.1103/PhysRevD.23.347}{\emph{Phys. Rev. D} {\bf 23} (1981) 347--356}.

\bibitem{Linde:1981mu}
A.~D. Linde, \emph{{A New Inflationary Universe Scenario: A Possible Solution of the Horizon, Flatness, Homogeneity, Isotropy and Primordial Monopole Problems}}, \href{http://dx.doi.org/10.1016/0370-2693(82)91219-9}{\emph{Phys. Lett. B} {\bf 108} (1982) 389--393}.

\bibitem{Starobinsky:1979ty}
A.~A. Starobinsky, \emph{{Spectrum of relict gravitational radiation and the early state of the universe}}, {\emph{JETP Lett.} {\bf 30} (1979) 682--685}.

\bibitem{Mukhanov:1981xt}
V.~F. Mukhanov and G.~V. Chibisov, \emph{{Quantum Fluctuations and a Nonsingular Universe}}, {\emph{JETP Lett.} {\bf 33} (1981) 532--535}.

\bibitem{Hawking:1982cz}
S.~W. Hawking, \emph{{The Development of Irregularities in a Single Bubble Inflationary Universe}}, \href{http://dx.doi.org/10.1016/0370-2693(82)90373-2}{\emph{Phys. Lett. B} {\bf 115} (1982) 295}.

\bibitem{Starobinsky:1982ee}
A.~A. Starobinsky, \emph{{Dynamics of Phase Transition in the New Inflationary Universe Scenario and Generation of Perturbations}}, \href{http://dx.doi.org/10.1016/0370-2693(82)90541-X}{\emph{Phys. Lett. B} {\bf 117} (1982) 175--178}.

\bibitem{Guth:1982ec}
A.~H. Guth and S.~Y. Pi, \emph{{Fluctuations in the New Inflationary Universe}}, \href{http://dx.doi.org/10.1103/PhysRevLett.49.1110}{\emph{Phys. Rev. Lett.} {\bf 49} (1982) 1110--1113}.

\bibitem{Bardeen:1983qw}
J.~M. Bardeen, P.~J. Steinhardt and M.~S. Turner, \emph{{Spontaneous Creation of Almost Scale - Free Density Perturbations in an Inflationary Universe}}, \href{http://dx.doi.org/10.1103/PhysRevD.28.679}{\emph{Phys. Rev. D} {\bf 28} (1983) 679}.

\bibitem{Betzios:2024oli}
P.~Betzios and O.~Papadoulaki, \emph{{Inflationary Cosmology from Anti-de Sitter Wormholes}}, \href{http://dx.doi.org/10.1103/PhysRevLett.133.021501}{\emph{Phys. Rev. Lett.} {\bf 133} (2024) 021501}, [\href{http://arxiv.org/abs/2403.17046}{{\tt 2403.17046}}].

\bibitem{Betzios:2024zhf}
P.~Betzios, I.~D. Gialamas and O.~Papadoulaki, \emph{{Magnetic anti{\textendash}de Sitter wormholes as seeds for Higgs inflation}}, \href{http://dx.doi.org/10.1103/9w85-fyhs}{\emph{Phys. Rev. D} {\bf 111} (2025) 123542}, [\href{http://arxiv.org/abs/2412.03639}{{\tt 2412.03639}}].

\bibitem{Betzios:2026rbv}
P.~Betzios, P.~Ghiringhelli, I.~D. Gialamas and O.~Papadoulaki, \emph{{A Menagerie of Wormholes and Cosmologies in the Gravitational Path Integral}},  \href{http://arxiv.org/abs/2602.23432}{{\tt 2602.23432}}.

\bibitem{Hartle:1983ai}
J.~B. Hartle and S.~W. Hawking, \emph{{Wave Function of the Universe}}, \href{http://dx.doi.org/10.1103/PhysRevD.28.2960}{\emph{Phys. Rev. D} {\bf 28} (1983) 2960--2975}.

\bibitem{Abdalla:2026mxn}
A.~I. Abdalla, S.~Antonini, R.~Bousso, L.~V. Iliesiu, A.~Levine and A.~Shahbazi-Moghaddam, \emph{{Consistent Evaluation of the No-Boundary Proposal}},  \href{http://arxiv.org/abs/2602.02682}{{\tt 2602.02682}}.

\bibitem{Betzios:2017krj}
P.~Betzios, N.~Gaddam and O.~Papadoulaki, \emph{{Antipodal correlation on the meron wormhole and a bang-crunch universe}}, \href{http://dx.doi.org/10.1103/PhysRevD.97.126006}{\emph{Phys. Rev. D} {\bf 97} (2018) 126006}, [\href{http://arxiv.org/abs/1711.03469}{{\tt 1711.03469}}].

\bibitem{Hebecker:2018ofv}
A.~Hebecker, T.~Mikhail and P.~Soler, \emph{{Euclidean wormholes, baby universes, and their impact on particle physics and cosmology}}, \href{http://dx.doi.org/10.3389/fspas.2018.00035}{\emph{Front. Astron. Space Sci.} {\bf 5} (2018) 35}, [\href{http://arxiv.org/abs/1807.00824}{{\tt 1807.00824}}].

\bibitem{Fu:2019oyc}
Z.~Fu and D.~Marolf, \emph{{Bag-of-gold spacetimes, Euclidean wormholes, and inflation from domain walls in AdS/CFT}}, \href{http://dx.doi.org/10.1007/JHEP11(2019)040}{\emph{JHEP} {\bf 11} (2019) 040}, [\href{http://arxiv.org/abs/1909.02505}{{\tt 1909.02505}}].

\bibitem{VanRaamsdonk:2021qgv}
M.~Van~Raamsdonk, \emph{{Cosmology from confinement?}}, \href{http://dx.doi.org/10.1007/JHEP03(2022)039}{\emph{JHEP} {\bf 03} (2022) 039}, [\href{http://arxiv.org/abs/2102.05057}{{\tt 2102.05057}}].

\bibitem{Antonini:2022blk}
S.~Antonini, P.~Simidzija, B.~Swingle and M.~Van~Raamsdonk, \emph{{Cosmology from the vacuum}}, \href{http://dx.doi.org/10.1088/1361-6382/ad1d46}{\emph{Class. Quant. Grav.} {\bf 41} (2024) 045008}, [\href{http://arxiv.org/abs/2203.11220}{{\tt 2203.11220}}].

\bibitem{Antonini:2022ptt}
S.~Antonini, P.~Simidzija, B.~Swingle and M.~Van~Raamsdonk, \emph{{Accelerating Cosmology from a Holographic Wormhole}}, \href{http://dx.doi.org/10.1103/PhysRevLett.130.221601}{\emph{Phys. Rev. Lett.} {\bf 130} (2023) 221601}, [\href{http://arxiv.org/abs/2206.14821}{{\tt 2206.14821}}].

\bibitem{Jonas:2023ipa}
C.~Jonas, G.~Lavrelashvili and J.-L. Lehners, \emph{{Zoo of axionic wormholes}}, \href{http://dx.doi.org/10.1103/PhysRevD.108.066012}{\emph{Phys. Rev. D} {\bf 108} (2023) 066012}, [\href{http://arxiv.org/abs/2306.11129}{{\tt 2306.11129}}].

\bibitem{Aguilar-Gutierrez:2023ril}
S.~E. Aguilar-Gutierrez, T.~Hertog, R.~Tielemans, J.~P. van~der Schaar and T.~Van~Riet, \emph{{Axion-de Sitter wormholes}}, \href{http://dx.doi.org/10.1007/JHEP11(2023)225}{\emph{JHEP} {\bf 11} (2023) 225}, [\href{http://arxiv.org/abs/2306.13951}{{\tt 2306.13951}}].

\bibitem{Antonini:2024bbm}
S.~Antonini and L.~G.~C. Bariuan, \emph{{Magnetic braneworlds: cosmology and wormholes}}, \href{http://dx.doi.org/10.1007/JHEP09(2024)070}{\emph{JHEP} {\bf 09} (2024) 070}, [\href{http://arxiv.org/abs/2405.18465}{{\tt 2405.18465}}].

\bibitem{Maloney:2025tnn}
A.~Maloney, V.~Meruliya and M.~Van~Raamsdonk, \emph{{Ordinary wormholes}},  \href{http://arxiv.org/abs/2503.12227}{{\tt 2503.12227}}.

\bibitem{Lavrelashvili:2026zsw}
G.~Lavrelashvili and J.-L. Lehners, \emph{{Nucleating an Inflationary Universe: Euclidean Wormholes and their No-Boundary Limit}},  \href{http://arxiv.org/abs/2603.11003}{{\tt 2603.11003}}.

\bibitem{Marolf:2021kjc}
D.~Marolf and J.~E. Santos, \emph{{AdS Euclidean wormholes}}, \href{http://dx.doi.org/10.1088/1361-6382/ac2cb7}{\emph{Class. Quant. Grav.} {\bf 38} (2021) 224002}, [\href{http://arxiv.org/abs/2101.08875}{{\tt 2101.08875}}].

\bibitem{Betzios:2025eev}
P.~Betzios, \emph{{A microscopic normal matrix model for (A)dS$_{2}$}}, \href{http://dx.doi.org/10.1007/JHEP01(2026)008}{\emph{JHEP} {\bf 01} (2026) 008}, [\href{http://arxiv.org/abs/2505.23891}{{\tt 2505.23891}}].

\bibitem{Cicoli:2023opf}
M.~Cicoli, J.~P. Conlon, A.~Maharana, S.~Parameswaran, F.~Quevedo and I.~Zavala, \emph{{String cosmology: From the early universe to today}}, \href{http://dx.doi.org/10.1016/j.physrep.2024.01.002}{\emph{Phys. Rept.} {\bf 1059} (2024) 1--155}, [\href{http://arxiv.org/abs/2303.04819}{{\tt 2303.04819}}].

\bibitem{McNamara:2020uza}
J.~McNamara and C.~Vafa, \emph{{Baby Universes, Holography, and the Swampland}},  \href{http://arxiv.org/abs/2004.06738}{{\tt 2004.06738}}.

\bibitem{Buttazzo:2013uya}
D.~Buttazzo, G.~Degrassi, P.~P. Giardino, G.~F. Giudice, F.~Sala, A.~Salvio et~al., \emph{{Investigating the near-criticality of the Higgs boson}}, \href{http://dx.doi.org/10.1007/JHEP12(2013)089}{\emph{JHEP} {\bf 12} (2013) 089}, [\href{http://arxiv.org/abs/1307.3536}{{\tt 1307.3536}}].

\bibitem{Bezrukov:2007ep}
F.~L. Bezrukov and M.~Shaposhnikov, \emph{{The Standard Model Higgs boson as the inflaton}}, \href{http://dx.doi.org/10.1016/j.physletb.2007.11.072}{\emph{Phys. Lett. B} {\bf 659} (2008) 703--706}, [\href{http://arxiv.org/abs/0710.3755}{{\tt 0710.3755}}].

\end{thebibliography}\endgroup

\end{document}